\documentclass[aps,preprint,prd,nofootinbib,superscriptaddress]{revtex4-1}	%superscriptaddress is an option for \author and \affiliation present
\pdfoutput=1	%arXiv required, within the first 5 lines
\usepackage[utf8]{inputenc}
\usepackage{booktabs}
\usepackage{amssymb,amsmath,amsfonts}
\usepackage{slashed}
\usepackage{indentfirst}
\usepackage[pdftex]{graphicx}
\usepackage{epstopdf}
\usepackage{psfrag}

\usepackage{float,multirow,longtable,rotfloat,supertabular} %for tables
\usepackage{hyperref}
\usepackage{mathrsfs}
\usepackage{url}
\usepackage{appendix}
\usepackage{subfigure}
\usepackage{color}
\usepackage[dvipsnames]{xcolor}
\usepackage{soul}
\usepackage{bm}
\usepackage{siunitx}
\usepackage{gensymb} %for degree symble

\DeclareUnicodeCharacter{2212}{-} %avoid LaTeX Error: Unicode character − (U+2212)

\allowdisplaybreaks	% allow long equation to display in two pages seperately in align environment

\newcommand{\GeV}{\,\text{GeV}}
\newcommand{\TeV}{\,\text{TeV}}

\graphicspath{{figures/}}

\def\figureautorefname~#1\null{Fig.\,#1\null}

\def\equationautorefname~#1\null{Eq.\,(#1)\null}

\newcommand{\La}{\mathcal{L}}
\newcommand{\Op}{\mathcal{O}}

\newcommand{\Afb}[1]{A^{0,#1}_{\rm FB}}

\newcommand{\eebb}{e^+e^-\to b\bar{b}}

\newcommand{\bpm}{\begin{pmatrix}}
\newcommand{\epm}{\end{pmatrix}}

\newcommand{\beq}{\begin{equation} }
\newcommand{\eeq}{\end{equation} }

\newcommand{\inab}{\,{\rm ab}^{-1}}

\begin{document}

\title{Tracing the bottom electroweak dipole operators at future lepton colliders}

\author{Jiayin Gu}
\email{jiayin\_gu@fudan.edu.cn}
\affiliation{Department of Physics and Center for Field Theory and Particle Physics, Fudan University, Shanghai 200438, China}
\affiliation{Key Laboratory of Nuclear Physics and Ion-beam Application (MOE), Fudan University, Shanghai 200433, China}

\author{Jiayu Guo}
\email{jiayuguo21@m.fudan.edu.cn}
\affiliation{Department of Physics and Center for Field Theory and Particle Physics, Fudan University, Shanghai 200438, China}

\author{Xiao-Ze Tan}
\email{xz\_tan@fudan.edu.cn}
\affiliation{Department of Physics and Center for Field Theory and Particle Physics, Fudan University, Shanghai 200438, China}
\affiliation{Deutsches Elektronen-Synchrotron DESY, Notkestr. 85, 22607 Hamburg, Germany}

\begin{abstract}

While often omitted in the SMEFT analyses of electroweak measurements, the electroweak dipole operators of the bottom quark have been found to be important in some cases and are also related to processes involving the top quark.  In this paper, we further investigate their effects, focusing on the measurements of the $\eebb$ process at a future lepton collider.  Their linear contributions are suppressed by the bottom mass due to the helicity flip in the interference term with the SM amplitude, leading to a nontrivial interplay between the linear and the quadratic contributions.  With two runs at the Z-pole and 240\,GeV, the effects of CP-even dipole coefficients can be well separated from the modifications of the SM $Zb\bar{b}$ couplings, while an additional run ({\it e.g.} at 360\,GeV) is useful for lifting a nontrivial second best-fit point due to the quadratic contributions.

\end{abstract}

\preprint{DESY-24-146} 

\maketitle

\tableofcontents

%%%%%%%%%%%%%%%%%%%%%%%%%%%%%%%
\section{Introduction}

The Standard Model (SM) of particle physics has been tested by collider experiments for several decades.  
In the electroweak (EW) sector, the SM predictions are in good agreement with the precision measurements carried out at the Large Electron-Positron collider (LEP) and Stanford Linear Collider (SLC) experiments~\cite{ALEPH:2005ab, ALEPH:2013dgf}, as well as those at the Large Hadron Collider (LHC). 
Higgs measurements have been performed in multiple channels at the LHC, and good precision has already been achieved in some channels with the accumulated data~\cite{ATLAS:2022vkf,CMS:2022dwd}. 
Several future lepton collider projects have also been proposed, including CEPC~\cite{CEPCStudyGroup:2018ghi, CEPCPhysicsStudyGroup:2022uwl}, FCC-ee~\cite{FCC:2018evy, Bernardi:2022hny}, ILC~\cite{ILCInternationalDevelopmentTeam:2022izu}, C$^3$~\cite{Bai:2021rdg}, CLIC~\cite{CLICdp:2018cto, CLIC:2018fvx}, as well as a possible high-energy muon collider~\cite{Aime:2022flm, Accettura:2023ked}. These future projects hold the potential to significantly enhance the current measurement precision of Higgs and EW observables, providing an even better probe of the SM and the possible new physics beyond it. 

Assuming the scale of the potential new physics is significantly larger than the EW scale, the Standard Model Effective Field Theory (SMEFT) provides a systematic and model-independent parameterization of the new physics effects, where deviations from the SM predictions (from new physics effects) are given by the Wilson coefficients of higher dimensional operators,
\beq
\La_{\rm SMEFT} = \La_{\rm SM} + \sum_{d>4} \frac{c^{(d)}_i}{\Lambda^{d-4}} \Op_i^{(d)} \,, \label{eq:Lsmeft}
\eeq
where $d$ is the dimension of the operator and $\Lambda$ parameterizes the scale of the potential underlying new physics.   
The dimension-six operators are expected to constitute the leading effects to the experimental observables (besides the dimension-5 operator which generates Majorana neutrino masses) and have been thoroughly studied. 
Many global analyses have been performed, with both the current measurements and the projected future ones~\cite{Falkowski:2015jaa, Durieux:2017rsg, Barklow:2017suo, Durieux:2018tev, Durieux:2018ggn, Ellis:2018gqa, DeBlas:2019qco, Durieux:2019rbz, Ellis:2020unq, Bruggisser:2022rhb,Liu:2022vgo, deBlas:2022ofj, Ethier:2021bye, Brivio:2022hrb, Bartocci:2023nvp, Allwicher:2023shc, Grunwald:2023nli, Garosi:2023yxg,Elmer:2023wtr,Kassabov:2023hbm,Bellafronte:2023amz,Celada:2024mcf,Asteriadis:2024qim}. 
In many cases, good overall precision can be achieved despite the large number of free parameters --- a clear demonstration of the usefulness of the global SMEFT framework. 

While the top dipole operators are considered in many SMEFT analyses at high-energy colliders
\cite{Buckley:2015lku,Schulze:2016qas,CMS:2019too,ATLAS:2019fwo,CMS:2021klw},
the dipole operators of other fermions are usually omitted, and for good reasons.\footnote{On the other hand, see Refs.~\cite{daSilvaAlmeida:2019cbr, Wen:2024cfu} on studies of light-quark dipoles at colliders and Ref.~\cite{Wen:2023xxc} on probing electron dipole operators with transverse spin asymmetry.
Refs.~\cite{Biswas:2021qaf, Biswas:2022fsr} also discussed the relevant dipole operators in the VBF Higgs and Higgs-photon production processes at the LHC.}
First, the SM amplitude and the amplitude with one dipole insertion have different helicity structures, and a mass insertion is needed to obtain an interference term.  For light fermions, this means that the new physics effects are strongly suppressed by fermion mass at the $1/\Lambda^2$ level and mainly given by the dipole-squared terms, which are at the $1/\Lambda^4$ level.  Second, low-energy electric or magnetic dipole measurements provide very stringent constraints on some of the dipole operators (especially the electron and muon ones~\cite{Muong-2:2006rrc,ACME:2018yjb,Muong-2:2008ebm,Muong-2:2021ojo}) which are far beyond the reaches of collider experiments.  However, for the second-heaviest fermion, the bottom quark, the above two reasons are not fully applicable, and the effects of dipole operators are indeed found to be important in some cases.   For instance, in Ref.~\cite{Liu:2022vgo}, it was found that the linear contributions of bottom dipole in EW observables, while suppressed by the bottom mass, tend to be more important than the loop effects of top operators, and the former should not be ignored if the latter is included in the analysis.  On the other hand, Ref.~\cite{Durieux:2019rbz} studied the quadratic contributions of bottom dipole operators (while setting the bottom mass to zero) and found their effects to be important in a global framework.   
Together, these two studies suggest that both the linear and the quadratic contributions of the bottom dipole operators could be important, and an analysis including the full contributions of bottom dipole operators is needed to understand their impacts.    

In this paper, we conduct a further investigation into the effects of bottom dipole operators in the $e^+e^- \to b\bar{b}$ process, with their full contributions (both linear and quadratic) taken into account.  
We focus on the scenario of a future circular lepton collider, with a tera-Z program that significantly improves the EW measurements at LEP/SLD, as well as runs at both 240\,GeV and around the $t\bar{t}$ threshold (around 360\,GeV).  
As a first step towards a more comprehensive global analysis, in this study, we focus on the CP-even two-bottom operators, including the dipole operators (assuming real Wilson coefficients) and those that modify the SM $Zb\bar{b}$ couplings.
We consider the rate (the total cross section, or $R_b$) and forward-backward asymmetry measurements, using the projected measurement reaches in recent Snowmass 2021 studies~\cite{CEPCPhysicsStudyGroup:2022uwl, deBlas:2022ofj}.   
Since the Z-pole measurements alone are not sufficient to simultaneously constrain all four parameters, a high energy run ({\it e.g.} at 240\,GeV) becomes crucial for the global analysis.  
Furthermore, as we will show, there is nontrivial interplay between the linear contributions of dim-6 operators and the dipole-squared contributions.  As a result, the fitting results with the Z-pole and 240\,GeV measurements exhibit a nontrivial second minimum due to cancellations from linear and quadratic contributions.  An additional run at a different energy ({\it e.g.} at 360\,GeV) helps to resolve this second minimum.

We also note that the bottom forward-backward asymmetry ($\Afb{b}$) measurement at LEP exhibits a $2.5\sigma$ deviation from the SM prediction~\cite{ALEPH:2005ab}.  Many studies have been carried out on the possible new physics explanations and future experimental tests of them (see {\it e.g.} Refs.~\cite{Choudhury:2001hs, He:2003qv, DaRold:2010as, Gori:2015nqa,Liu:2017xmc,Yan:2021veo,Yan:2021htf,Breso-Pla:2021qoe,Chai:2022ued,Dong:2022ayy,Yan:2023ccj,Bishara:2023qhe}). 
A non-zero bottom dipole interaction would also modify the SM prediction for $\Afb{b}$. In particular, the dipole-squared contribution tends to reduce the asymmetry which is in the direction that the LEP measurement prefers.  
However, we find that a large coefficient ($|\kappa_{bZ}|\sim 0.001$, defined in \autoref{eq:La}) is needed for a good fit with the LEP measurement.  Assuming dipole interactions are generated at the one-loop level, this would require the new physics scale to be very low, likely out of the perturbative SMEFT regime.

The rest of this paper is organized as follows. In \autoref{sec:theory}, we introduce the SMEFT framework as well as the effective couplings for our analysis. 
The details of the analysis are described in \autoref{sec:ana}.  
Our results in terms of the global fit with four effective couplings are presented in \autoref{sec:results}, including the preferred regions in the parameter space and the numerical values of the 1-sigma bounds and correlations.  
Finally, we summarize our work in \autoref{sec:summary}.
In Appendix~\ref{app:dim6}, we provide the results in terms of the dim-6 Wilson coefficients. In Appendix~\ref{app:runningmass}, we provide additional results which use a different bottom mass value (with the renormalization scale $\mu=m_Z$ instead of $\mu=m_b$).

%%%%%%%%%%%%%%%%%%%%%%%%%%%%%%%
\section{The SMEFT framework}
\label{sec:theory}

We focus on the CP-even two-bottom dimension-6 operators in our study, which are (as in {\it e.g.} the Warsaw basis~\cite{Grzadkowski:2010es})  
\begin{align}
	\mathcal{O}_{H q}^{(1)} =&~ \left ( H^\dagger i \overleftrightarrow{D_{\mu } } H  \right ) (\bar{q}_L \gamma^{\mu }q_L ) \,, \nonumber\\   
	\mathcal{O}_{H q}^{(3)} =&~  \left ( H^\dagger i \overleftrightarrow{D_{\mu }^{i} } H  \right ) (\bar{q}_L \sigma^i \gamma^{\mu }q_L )   \,, \nonumber\\
	\mathcal{O}_{H b} =&~ \left ( H^\dagger i \overleftrightarrow{D_{\mu }} H  \right ) (\bar{b}_R \gamma^{\mu }b_R)      \,, \nonumber\\
	\mathcal{O}_{b W} =&~ \left(\bar{q}_L \sigma^{\mu \nu} b_R \right) \sigma^{i} H W_{\mu \nu}^{i}    \,, \nonumber\\
	\mathcal{O}_{b B} =&~ \left(\bar{q}_L \sigma^{\mu \nu} b_R \right) H B_{\mu \nu}    \,, \label{eq:bop}
\end{align}
where $q_L \equiv \bpm t_L \\ b_L \epm$ is the third generation quark doublet.\footnote{In a more general framework, these operators can have more complex flavor structures.  Here we consider only the third generation quarks.}  The dipole operators are not hermitian and can have complex coefficients in general.  Here we focus on CP-even effects and assume that the dipole coefficients are real. The imaginary parts of dipole operators, which give raises to CP-violation effects, are stringently constrained by the measurements of the electron electric dipole moment (EDM). For instance, Ref.~\cite{Panico:2018hal} considered the two-loop effects on the electron EDM in SMEFT and found that the size of the coefficients Im[$c_{bW}$] and Im[$c_{bB}$] are constrained to be less than one, assuming $\Lambda =10\TeV$, 
based on the measurement by the ACME collaboration in 2018~\cite{ACME:2018yjb}.  These constraints are likely to be further improved by the next generation ACME experiment.

After electroweak symmetry breaking (EWSB), we have the following interactions between the bottom quark and the EW gauge bosons 
\beq
\La \supset - e A_\mu \bar{b} \gamma^\mu b - \frac{g}{\cos{\theta_W}} Z_\mu\left( g_{L b} \bar{b}_L \gamma^\mu b_L+ g_{R b} \bar{b}_R \gamma^\mu b_R\right) +  
\frac{\kappa_{bA}}{m_b} \left(\bar{b} \sigma^{\mu\nu} b \right) A_{\mu\nu} + \frac{\kappa_{bZ}}{m_b} \left(\bar{b} \sigma^{\mu\nu} b \right) Z_{\mu\nu}  \,, \label{eq:La}
\eeq
where $e$ is the electric coupling, $g$ is the weak coupling, $\theta_W$ is the weak mixing angle, and we have normalized the (CP-even) dipole couplings $\kappa_{bA}$ and $\kappa_{bZ}$ by $m_b$ to make them dimensionless.  
The couplings $g_{L b}$ and $g_{R b}$ contain both the SM and new physics contributions, 
\beq
 g_{L b}= -\frac{1}{2} + \frac{1}{3} \sin^2\theta_W +\delta g_{L b}, \hspace{1cm} g_{R b}=\frac{1}{3} \sin^2\theta_W +\delta g_{R b} \,,
\eeq
where the new physics contributions $\delta g_{L b}$ and $\delta g_{R b}$ are related to the coefficients of the first three operators in \autoref{eq:bop} (the convention of the Wilson coefficients follows the one in \autoref{eq:Lsmeft})
\beq
	\delta g_{Lb}=(c_{H q}^{(1)}+ c_{H q}^{(3)})  \frac{v^{2} }{2 \Lambda^{2} } \,, \hspace{1cm}
	\delta g_{Rb}=c_{H b}
	\frac{v^{2} }{ 2\Lambda^{2} } \,,
\eeq
while the dipole couplings are given by 
\begin{align}
		\kappa_{bZ} =&~ \frac{m_b v}{\sqrt{2} \Lambda^{2}}\left(\cos\theta_W \, c_{b W}+\sin\theta_W \, c_{b B}\right) \,, \nonumber\\ 
		\kappa_{bA} =&~ \frac{m_b v}{\sqrt{2} \Lambda^{2}}\left(\cos\theta_W \, c_{b B}-\sin\theta_W \, c_{b W}\right) \,.  \label{eq:kappabZA}
\end{align}

We use the $\{m_Z, m_W, \, G_F \}$ input scheme with the following values for the input parameters~\cite{ParticleDataGroup:2024cfk}:
\beq
m_Z \approx 91.18\,{\rm GeV} \,, \hspace{1cm} 
m_W \approx 80.37\,{\rm GeV} \,, \hspace{1cm} 
G_{F} \approx 1.166\times10^{-5}\, \text{GeV}^{-2} \,. 
\eeq
The choice of EW input scheme has little impact on our study, since we focus on the bottom measurements.  
On the other hand, the choice of the bottom mass has a significant impact, since it determines the size of the interference term between the SM and the dipole contribution.  
We use the value in the $\overline{\rm MS}$ scheme with the renormalization scale $\mu=m_b$, which gives $m_b \approx 4.18$\,GeV~\cite{ParticleDataGroup:2024cfk}.  
The reason to choose $\mu=m_b$ instead of $\mu=m_Z$ is that here the bottom mass characterizes the helicity flip of the bottom quark, so that a definition closer to the pole mass is more appropriate.  This is different from the treatment of the bottom mass in the $h\to b\bar{b}$ process, which needs to be run to $\mu=m_h$.   

For the latter, the running of the bottom mass (or more precisely, the bottom Yukawa coupling) effectively resums the leading logs of the higher-order QCD corrections, which is important for obserables such as $\Gamma(h\to b\bar{b})$.   On the other hand, the observables $R_b$ and $\Afb{b}$ are ratios of hadronic observables where one might expect large cancellations of QCD corrections.  A proper treatment of the higher-order QCD corrections is also beyond the scope of our study.    
For comparison, we also show in Appendix~\ref{app:runningmass} the fit results that use $\mu=m_Z$ instead of $\mu=m_b$.   
We note that the choice of bottom mass has also been discussed in previous studies (see {\it e.g.} Refs.~\cite{Catani:1999nf,Bernreuther:2016ccf,deBlas:2024bmz}) but in the context of SM predictions for the Z-pole observables (such as $\Afb{b}$), for which the impacts are relatively small.

%%%%%%%%%%%%%%%%%%%%%%%%%%%%%%%%%%%%
\section{Measurements and Analyses}
\label{sec:ana}

For the future $e^+e^-$ collider, we use the most recent run scenario of the CEPC from the Snowmass study~\cite{CEPCPhysicsStudyGroup:2022uwl}, with an integrated luminosity of $100\inab$ at the Z-pole, $20\inab$ at 240\,GeV, and $1\inab$ at 360\,GeV.  CEPC is also expected to collect $6 \inab$ data at the $WW$ threshold.  However, as mentioned later, we do not include the $WW$ threshold run at the moment due to the lack of measurement projections.  
Given the similarity between the CEPC and FCC-ee run scenarios, we expect similar results from our analysis if FCC-ee inputs are used instead.   

Given that we focus on the bottom operators, we also consider only the observables that are directly related to these operators.  At the Z-pole, the main observables are $R_b$, $\Afb{b}$ and $A_b$, which are summarized in \autoref{table:zpoledata}.\footnote{While other observables, such as the total hadronic cross section at Z-pole, $\sigma_{had}^0$, also receive contributions from the bottom operators, they are also sensitive to many other operators, and are thus not included in our analysis.}  The current measured values (from LEP and SLD) and the SM predictions are from the PDG~\cite{ParticleDataGroup:2024cfk} (see also Ref.~\cite{ALEPH:2005ab}), while the projected CEPC reach are from Ref.~\cite{CEPCPhysicsStudyGroup:2022uwl}.  Note that, without beam polarization, CEPC could not directly measure $A_b$.  However, the projection of $A_b$ is provided in Ref.~\cite{CEPCPhysicsStudyGroup:2022uwl}, which is derived from the measurements of $\Afb{b}$ and $A_e$, while the projection of $\Afb{b}$ is not provided.  We therefore directly use the projection of $A_b$ reported in Ref.~\cite{CEPCPhysicsStudyGroup:2022uwl}.  
For measurements at higher energies, there are no official CEPC projections.
We use the projections in Ref.~\cite{deBlas:2022ofj} on the total cross section of $\eebb$ and $\Afb{b}$ at 240\,GeV and 360\,GeV which are based on simple estimations with only statistical uncertainties. In particular, an overall universal efficiency of $15\%$ is applied to signal events, while $\cos\theta$ (where $\theta$ is the production polar angle) is assumed to be within the range $[-0.9,\,0.9]$.  The projections are summarized in \autoref{table:CEPC240360data}.  While the WW threshold run is potentially relevant for our study, there is no corresponding projections from Ref.~\cite{deBlas:2022ofj}, and we do not include it in the analysis.  
For all future measurements, we assume that the central values are SM-like.

\begin{table}[t]
    \centering
    \setlength{\tabcolsep}{0.5em}{
    \begin{tabular}{cccc}
   \hline \text { Quantity } & \text { LEP/SLD measurement } & \text { SM prediction } & Projected CEPC precision \\ \hline
$R_{b}$ & $0.21629 \pm 0.00066$ & $0.21582 $ & $ 4.4 \times 10^{-5} $ \\
$\Afb{b}$ & $0.0996 \pm 0.0016$ & $0.1029 $ & - \\
$A_{b}$ & $0.923 \pm 0.020$ & 0.9347 & $ 2.1 \times 10^{-4} $ \\
   \hline
    \end{tabular} }
    \caption{The current results and future projections of Z-pole measurements, obtained from \cite{ParticleDataGroup:2024cfk,CEPCPhysicsStudyGroup:2022uwl}. Note that the CEPC will measure $\Afb{b}$ but Ref.~\cite{CEPCPhysicsStudyGroup:2022uwl} only reports the projection for $A_b$ which is derived from the measurement of $\Afb{b}$ and $A_e$. 
 }
    \label{table:zpoledata}
\end{table}

\begin{table}[t]
    \centering
    \setlength{\tabcolsep}{0.5em}{
    \begin{tabular}{ccccc}
      \hline
      $\sqrt{s}[\mathrm{GeV}] $ &   $\mathcal{L}\left[\mathrm{ab}^{-1}\right] $ & $ \sigma(\eebb)[\mathrm{fb}]$  &  $\Afb{b}$   \\
   \hline
      240 &    20 &  275.64 $\pm$ 0.12  &  0.592 $\pm$ 0.00034  \\
       360 &  1 &  108.33 $\pm$ 0.33  &  0.602 $\pm$ 0.0024   \\
 \hline 
    \end{tabular} }
    \caption{The projected $\eebb$ measurements at  CEPC from Ref.~\cite{deBlas:2022ofj}.  Only the signal statistical uncertainties are considered, with a universal selection efficiency of $\epsilon =0.15$.  The production polar angle is assumed to be in the range $-0.9 <\cos\theta<0.9$.}
    \label{table:CEPC240360data}   
\end{table}

We compute all observables at the tree level in terms of the four effective parameters $\left\{ \delta g_{Lb} ,\, \delta g_{Rb} ,\, \kappa_{bA} ,\,  \kappa_{bZ} \right\}$ described in the previous section.  To be consistent with the SM predictions in \autoref{table:zpoledata} which include loop corrections, the SMEFT predictions are multiplied by the ratio between SM full prediction and SM tree-level prediction, when necessary. 
Two scenarios are considered:  In the first one, the observables are expanded in terms of the effective parameters and only the linear contributions are kept.  In the second one, the full expression is used without any expansion.\footnote{This case does not necessarily corresponds to a quadratic relation since $R_b$, $\Afb{b}$ and $A_b$ are ratios in which the effective parameters contribute to both the numerator and the denominator.  Nevertheless, the numerical difference is small beyond the square terms.}

For simplicity, we assume the measurements are uncorrelated, in which case the total chi-square is constructed as
\beq
\chi^2 = \sum_i \frac{\left( O_{i,\text{exp}}-O_{i,\text{th}}\right)^2}{\left( \delta O_{i} \right)^2} \,,
\eeq
where $O_{i,\text{th}}$, $O_{i,\text{exp}}$, $\delta O_{i}$ are the theory prediction, the measured central value and the one-sigma uncertainty for the observable $O_i$, respectively.  In the case where the observables depend linearly on the effective parameters (denoted as $c_i$ below), one could write %the $\chi^2$ can be written as
\beq
\Delta\chi^2 \equiv  \chi^2 - \chi^2_{\rm min}=\sum_{i j}\left(c_i-c_{0,i}\right)\left[\sigma^{-2}\right]_{i j}\left(c_j-c_{0,j}\right) \,, \label{eq:chis2}
\eeq
where $c_{0,i}$ are the best fitted values and $\sigma^{-2} = \left(  \boldsymbol{\delta}\boldsymbol{c}^T \, \boldsymbol{\rho} \, \boldsymbol{\delta} \boldsymbol{c}\right)^{-1}$ is the inverse covariance matrix.  The one-sigma precision $\delta c_i$ and the correlation matrix $\rho_{ij}$ can be obtained from $\sigma^{-2}$.  Note that, since we assume all central values are SM-like for future measurements,  $\chi^2_{\rm min}$ and the central values $c_{0,i}$ are zero by construction.

%%%%%%%%%%%%%%%%%%%%%%%%%%%%%%%%%%%%
\section{Results}
\label{sec:results}

With the projected CEPC measurements in \autoref{table:zpoledata} and \autoref{table:CEPC240360data}, we perform a global analysis with the four parameters $\left\{ \delta g_{Lb} ,\, \delta g_{Rb} ,\, \kappa_{bA} ,\,  \kappa_{bZ} \right\}$.  All the measurement central values are assumed to be SM-like.  One important observation is that the Z-pole measurements alone provide only two independent constraints in terms of the four parameters.  On the other hand, the dipole couplings have different chiral structures and energy dependence.  
As such, the measurements at high energies are crucial for separating the effects of the dipole couplings from the SM gauge couplings.  
The ratio between the SM $Z$ and $\gamma$ diagrams also changes with respect to the center-of-mass energy, so having measurements at different energies helps to  distinguish the dipole coupling to $Z$ ($\kappa_{bZ}$) and the one to photon ($\kappa_{bA}$).  

\begin{figure}[t]
	\centering
	
	\subfigure[ ~keeping only the linear dependence of $\left\{ \delta g_{Lb} ,\, \delta g_{Rb} ,\, \kappa_{bA} ,\,  \kappa_{bZ} \right\}$ in the observables ]{
		
		\includegraphics[width=0.45\textwidth]{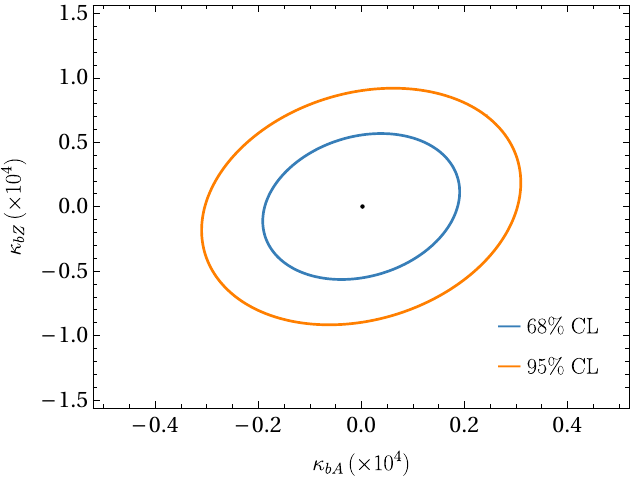}
	\hspace{0.2cm}
		
		\includegraphics[width=0.45\textwidth]{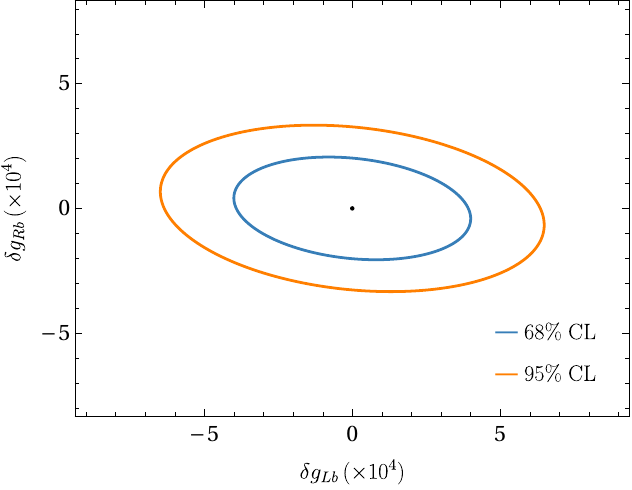}}

	\subfigure[ ~keeping the full dependence of $\left\{ \delta g_{Lb} ,\, \delta g_{Rb} ,\, \kappa_{bA} ,\,  \kappa_{bZ} \right\}$ ]{
		
		\includegraphics[width=0.45\textwidth]{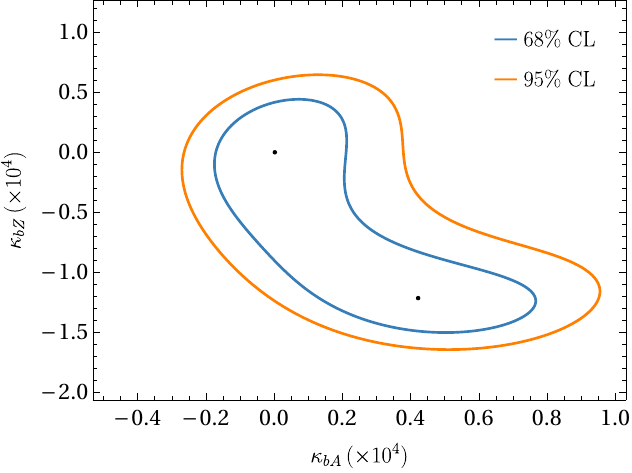}
	\hspace{0.2cm}
		
		\includegraphics[width=0.45\textwidth]{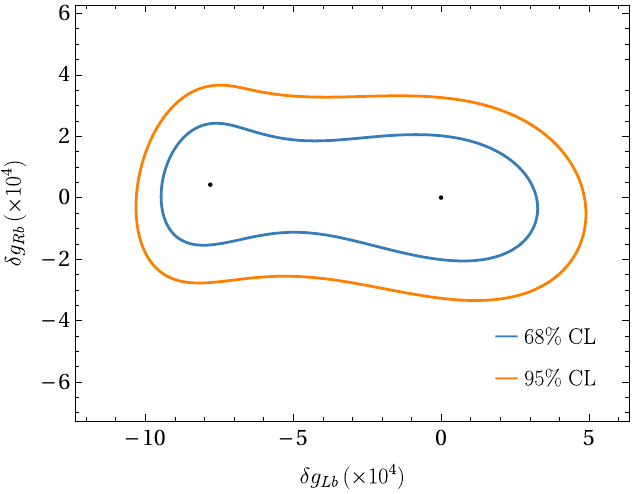}}  
	
	\caption{%
    The preferred region from a global fit with $ \delta g_{Lb} $, $ \delta g_{Rb} $, $\kappa_{bA} $, $ \kappa_{bZ} $ to the combination of the future Z-pole and $\sqrt{s}= 240\GeV$ measurements. 
 The results in the top (bottom) panel corresponds to the linear (full) fit.  The black dots correspond to $\chi^2=\chi^2_{\rm min}=0$.  The second minimum (non-SM best fit point) of the full fit is at $\left\{ \delta g_{Lb} ,\, \delta g_{Rb} ,\, \kappa_{bA} ,\,  \kappa_{bZ} \right\} = \left\{ -7.80,0.420,0.423,-1.22 \right\} \times 10^{-4}$.
    }
	\label{fig:zpole240}
\end{figure}
In \autoref{fig:zpole240}, we present the preferred regions in the parameter space of $\left\{ \delta g_{Lb} ,\, \delta g_{Rb} ,\, \kappa_{bA} ,\,  \kappa_{bZ} \right\}$, obtained from a four-parameter global fit to the measurements at the Z-pole and 240\,GeV.  The results in the top panel are obtained by keeping only the linear dependence of the effective parameters in the observables, while for the bottom panel the full dependence is kept.  For later convenience, we will refer to the two scenarios by ``linear fit'' and ``full fit'', respectively.  In each case, the contours at the confidence level of 68\% and 95\% are projected on the $(\kappa_{bA}, \kappa_{bZ})$ plane (left panel) and the $(\delta g_{Lb}, \delta g_{Rb})$ plane (right panel) by profiling over the other two parameters.  The measurements at two different energies provide four independent constraints on the four parameters, and from the results we see that indeed the four parameters can be constrained simultaneously.  
On the other hand, we observe a visible difference between the results of the linear fit and the full fit, suggesting that the squared terms of the dipole contributions play an important role in the fit.  The full fit results also exhibit a peculiar second best-fit point, where all the contributions from the four parameters $\left\{ \delta g_{Lb} ,\, \delta g_{Rb} ,\, \kappa_{bA} ,\,  \kappa_{bZ} \right\}$ cancel in the four observables, which takes the SM values.  Such points are indeed possible since setting the four observables to SM values gives a set of non-linear equations that may have multiple solutions.  Overall, the precision of the effective parameters is at the $10^{-4}$ level or better.

\begin{figure}[t]

	\subfigure[~keeping only the linear dependence of $\left\{ \delta g_{Lb} ,\, \delta g_{Rb} ,\, \kappa_{bA} ,\,  \kappa_{bZ} \right\}$ in the observables ]{
		
		\includegraphics[width=0.48\textwidth]{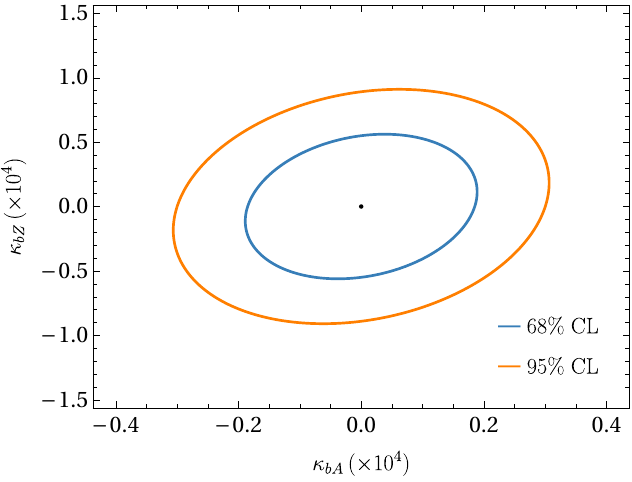}
    \hspace{0.2cm}
		\includegraphics[width=0.48\textwidth]{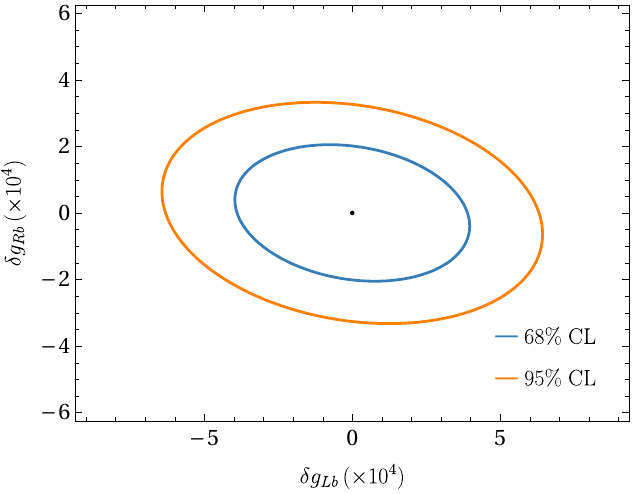}}

	\centering
	\subfigure[~keeping the full dependence of $\left\{ \delta g_{Lb} ,\, \delta g_{Rb} ,\, \kappa_{bA} ,\,  \kappa_{bZ} \right\}$ ]{
		
		\includegraphics[width=0.48\textwidth]{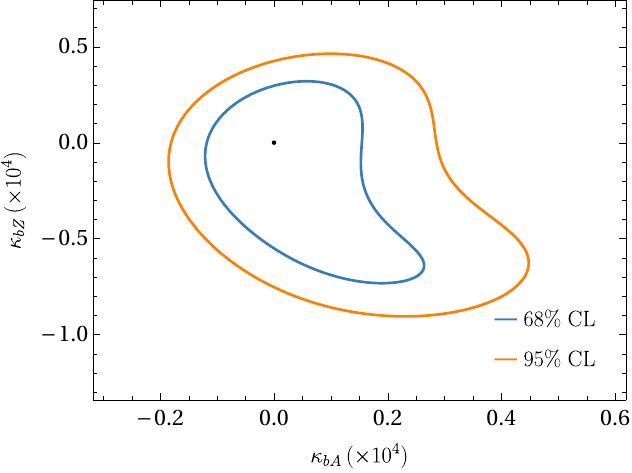}
    \hspace{0.2cm}
		\includegraphics[width=0.48\textwidth]{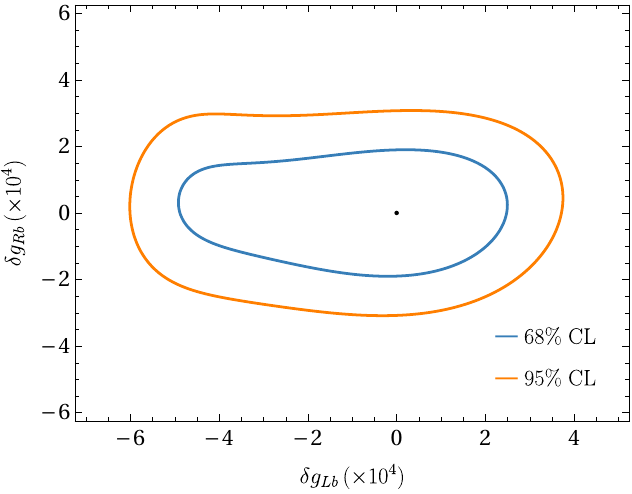}}
	
	\caption{%
    Same as \autoref{fig:zpole240} but also includes the 360\,GeV measurements.  The second minimum in the full fit is lifted. 
     }
	\label{fig:zpole240360}
\end{figure}

In \autoref{fig:zpole240360}, the same results are presented but also with the 360\,GeV run.  Due to the relatively low luminosity and also smaller SM cross sections at higher energies, the measurements at 360\,GeV has larger uncertainties, as shown in \autoref{table:CEPC240360data}.  For the linear fit, 
the addition of the 360\,GeV measurements provides only marginal improvements on the results.  However, the 360\,GeV measurements are crucial for the full fit, and can help to lift the ``second minimum'' exhibited by the results in \autoref{fig:zpole240}.  We also observe a sizable difference between \autoref{fig:zpole240} and \autoref{fig:zpole240360} for the full fit results. 
Note that the WW threshold run could also serve as the run to lift the ``second minimum'', which would be crucial if the 360\,GeV run is not available.  

\begin{table}[t]
    \centering
	\renewcommand\arraystretch{1.0}
	\vspace{0.5em}\centering
	\setlength{\tabcolsep}{0.5em}{
    \begin{tabular}{cccccc|c}
    \multicolumn{7}{c}{Z-pole + 240\,GeV } \\
    \hline 
                & Linear fit                             & \multicolumn{4}{c|}{Correlation $\rho$}                   & Full fit                            \\
                & $1 \sigma$ bound ($\times 10^{-4}$) & $\delta g_{Lb}$ & $\delta g_{Rb}$ & $\kappa_{bZ}$ & $\kappa_{bA}$ & $1 \sigma$ bound ($\times 10^{-4}$) \\ \hline
    $\delta g_{Lb}$ & $\pm 2.65$                          & 1               &                 &               &               & [-8.90, 2.20]                   \\
    $\delta g_{Rb}$ & $\pm 1.36$                          & -0.200           & 1               &               &               & [-1.33, 1.68]                   \\
    $\kappa_{bZ}$   & $\pm 0.375$                         & 0.975           & -0.271           & 1             &               & [-1.408, 0.304]                  \\
    $\kappa_{bA}$   & $\pm 0.127$                         & 0.112           & -0.088           & 0.201         & 1             & [-0.118, 0.655]                 \\ \hline \vspace{0cm}
    \end{tabular}  }
	\setlength{\tabcolsep}{0.5em}{
    \begin{tabular}{cccccc|c}
        \multicolumn{7}{c}{Z-pole + 240\,GeV + 360\,GeV } \\
        \hline 
               & Linear fit                             & \multicolumn{4}{c|}{ Correlation $\rho$}                   & Full fit                            \\
                & $1 \sigma$ bound ($\times 10^{-4}$) & $\delta g_{Lb}$ & $\delta g_{Rb}$ & $\kappa_{bZ}$ & $\kappa_{bA}$ & $1 \sigma$ bound ($\times 10^{-4}$) \\ \hline
    $\delta g_{Lb}$ & $\pm 2.63$                          & 1               &                 &               &               & [-2.88, 1.68]                   \\
    $\delta g_{Rb}$ & $\pm 1.36$                          & -0.193           & 1               &               &               & [-1.20, 1.20]                   \\
    $\kappa_{bZ}$   & $\pm 0.372$                         & 0.975           & -0.264           & 1             &               & [-0.42, 0.225]                \\
    $\kappa_{bA}$   & $\pm 0.125$                         & 0.112           & -0.088          & 0.202         & 1             & [-0.078, 0.096]                 \\ \hline
    \end{tabular}   }   
    	\caption[results]{1$ \sigma $ bound and correlations of $ \delta g_{Lb} $, $ \delta g_{Rb} $, $ \kappa_{bZ} $, $ \kappa_{bA} $ with expected measurements of EW observables at CEPC.  The results in the top panel are obtained from the combination of Z-pole and 240\,GeV measurements, while for the results in the bottom panel the 360\,GeV measurements are also included.  The shown correlation matrices correspond to the linear fits. 
     }
    \label{tab:fit1}
\end{table}

In \autoref{tab:fit1}, we present the numerical results of the 4-parameter global fit.  For the linear fit, the one-sigma bounds and the correlation matrices are shown, which are obtained directly from \autoref{eq:chis2}.  For the full fit, we present the one-sigma bounds obtained by profiling all other parameters.  One interesting observation here is that $\kappa_{bA}$ is generally better constrained than $\kappa_{bZ}$.  This is because $\kappa_{bA}$ contributes to the photon diagram, and can thus be more easily distinguished from the other couplings that contribute to the $Z$ diagram.  On the other hand, it is harder to distinguish $\kappa_{bZ}$ from $\delta g_{Lb}$ and $\delta g_{Rb}$.  This is reflected in the correlation matrices in \autoref{tab:fit1}.

In Appendix~\ref{app:dim6}, we also present the same results in terms of the dim-6 Wilson coefficients instead of the effective parameters.  Setting $\Lambda=1$\,TeV, the Wilson coefficients are constrained to the $\mathcal{O}(10^{-2} \sim 10^{-3})$ level.  This provides a confirmation on the validity of the EFT expansion since the collider energy is at $\sim 10^2$\,GeV.

%%%%%%%%%%%%%%%%%%%%%%%%%%%%%%%%%%%%
\section{Summary}
\label{sec:summary}

In this paper, we performed a phenomenological analysis on the $\eebb$ measurements at a future lepton collider to extract the precision reach on a set of dimension-6 bottom-quark operators.  Focusing on CP-even effects, we found that the two bottom dipole operators, $\mathcal{O}_{b W} = \left(\bar{q}_L \sigma^{\mu \nu} b_R \right) \sigma^{i} H W_{\mu \nu}^{i} $ and $\mathcal{O}_{b B} = \left(\bar{q}_L \sigma^{\mu \nu} b_R \right) H B_{\mu \nu}$, could have sizable contributions to the $\eebb$ process despite the suppression of the interference term with SM by the bottom mass.  The combination of measurements at multiple center-of-mass energies is crucial for discriminating dipole operators from the ones that modify the SM $Zb\bar{b}$ couplings.  
The dipole quadratic contributions (from the square of the dipole amplitudes) are not suppressed by the bottom mass, and their effects can be comparable with the linear ones.  With measurements at only two energies ({\it e.g.} Z-pole and 240\,GeV), the inclusion of the quadratic contributions leads to a non-SM second best-fit point in the parameter space due to cancellations among the linear and quadratic terms, which can be lifted by including the measurements at an additional energy ({\it e.g.} 360\,GeV).  Overall, we found that a future lepton collider, with measurements at both the Z-pole and higher energies, provides powerful probes on the bottom dipole operators, with a precision reach corresponding to a new physics scale in the multi-TeV range, assuming an order-one coupling.

The importance of the dipole quadratic contributions in our analysis naturally raises concerns about the validity of the SMEFT framework, since they are at the $1/\Lambda^4$ level, formally the same as the linear contribution of dimension-8 operators.  It should be noted that, for dimension-8 dipole operators, the interference term with the SM is also suppressed by the bottom mass.  Treating the bottom mass as another small expansion parameter, one could indeed keep the quadratic contribution of dimension-6 dipole operators while omitting the linear dimension-8 ones.  
We also note that the Wilson coefficients that modify the SM $Zb\bar{b}$ couplings, $c^{(1)}_{Hq}+c^{(3)}_{Hq}$ and $c_{Hb}$, are much better constrained than the dipole coefficients, as shown in \autoref{tab:fit2}, and their square terms have little impact in our analysis.  As such, we do not expect that the dimension-8 operators that modify the SM $Zb\bar{b}$ couplings are relevant to our analysis. It is indeed nontrivial to build a model that generates the bottom dipole interactions while keeping modifications to the SM $Zb\bar{b}$ couplings small. This is however not a major concern in our study, since we take a bottom-up approach assuming Wilson coefficients are free parameters subject to experimental constraints, as in many other SMEFT analyses.
More generally, the robustness of the SMEFT analysis can be improved by either cutting the phase space where the quadratic contributions dominate~\cite{Contino:2016jqw}, or treating the quadratic contributions as the theory error~\cite{Alte:2017pme}.  Here we chose to simply present both the results with the quadratic contributions and the ones without, leaving a more delicate treatment of SMEFT validity or the construction of viable UV models to future studies. We also note that the differences between the linear-fit results and the full-fit ones in \autoref{tab:fit1} are relatively small (with at most $\sim 40\%$ relative differences) if all measurements up to 360\,GeV are included.  Therefore, we would expect the results to be still within the same order of magnitude, should a more strict treatment of EFT validity be applied ({\it e.g.} by treating the quadratic contributions as theory error).

There are several directions in which our study can be extended.  We restricted the collider scenario to a circular collider without beam polarizations.  It is straightforward to apply the same analysis to linear colliders with beam polarizations.  
In addition to the possibility of directly measuring the asymmetry observable $A_b$, with different beam polarizations one could also effectively change the ratio between the $Z$ and photon diagrams, which receive different dipole contributions.  
As such, beam polarization could provide additional discrimination power in the fit.  
We have focused on the CP-even two-bottom operators in our study, and an obvious extension is to also consider the CP-odd effects.  This introduces more parameters, and it could be challenging to find observables that separate the CP-odd effects from the CP-even ones. \footnote{This is unlike the top quark, for which one could exploit the top decay distribution.}  It is also important to implement our analysis in a more global framework, in particular with the top-quark operators, as already pointed out in \cite{Durieux:2019rbz, Liu:2022vgo}.  We leave these important directions to future studies.

%%%%%%%%%%%%%%%%%%%%%%%%%%%%%%%%%%%%
\section*{Acknowledgments }

The authors thank Yong Du, Christophe Grojean, Yiming Liu, Bin Yan for useful discussions and valuable comments on the manuscripts. X.Z. Tan also thanks Lukas Allwicher, Lisong Chen for discussions about quark running mass, and Guilherme Guedes especially for help on the UV model searching with the SOLD package although not being included ultimately in this work.
This work is supported by the National Natural Science Foundation of China (NSFC) under grant No. 12035008, No. 12375091 and No. 12347171.
X.Z. Tan also acknowledges the support from Helmholtz – OCPC (Office of China Postdoctoral Council) Postdoctoral Fellowship Program.

%%%%%%%%%%%%%%%%%%%%%%%%%%%%%%%%%%%%

\appendix

\section{Results in terms of the dim-6 Wilson coefficients}
\label{app:dim6}

%%%%%%%%%%%%%%%%%%%%%%%%%%%%%%%%%%%%

\begin{figure}[htb]
	\centering
 
    \subfigure[~keeping only the linear dependence of Wilson coefficients]{
		
		\includegraphics[width=0.45\textwidth]{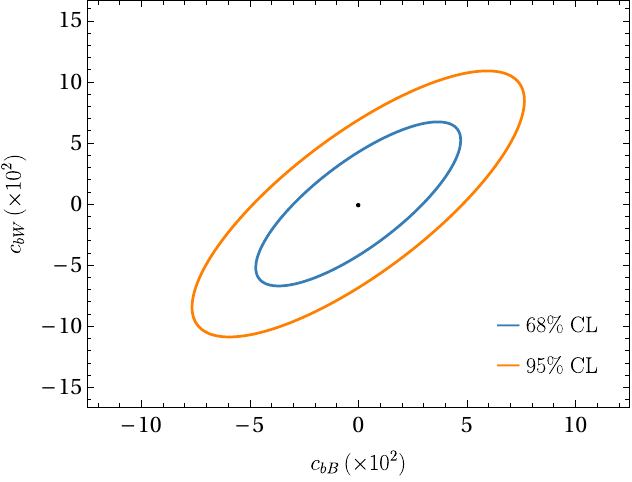}
	\hspace{0.2cm}
		\includegraphics[width=0.45\textwidth]{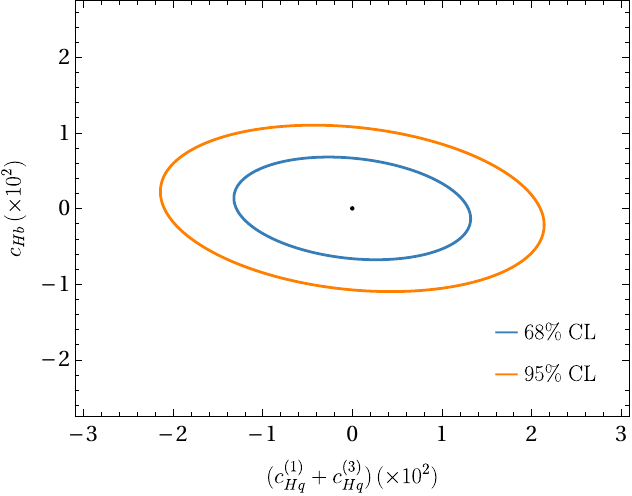}}
        
	\subfigure[~keeping the full dependence of Wilson coefficients]{
		
		\includegraphics[width=0.45\textwidth]{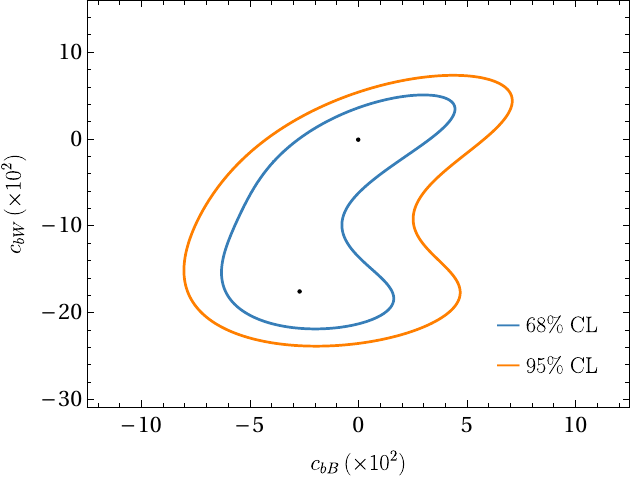}
	\hspace{0.2cm}
		\includegraphics[width=0.45\textwidth]{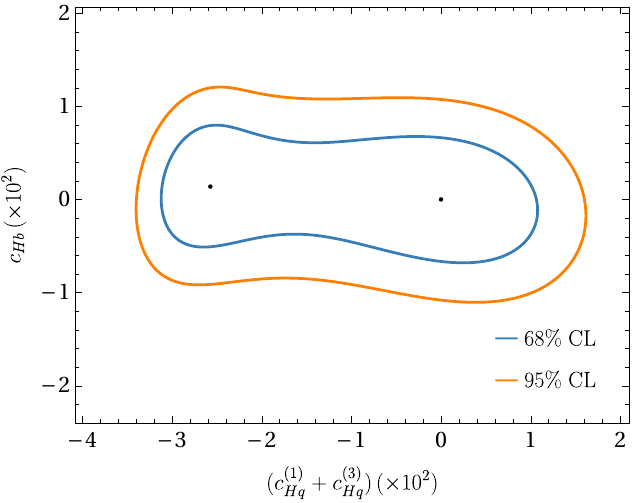}}
	
	\caption{%
    The preferred region from a global fit with $c_{Hq}^{(1)}+c_{Hq}^{(3)}$ , $c_{Hb} $ ,  $ c_{bB}$  and  $c_{bW} $ to the combination of the Z-pole and $\sqrt{s}= 240\GeV$ measurements. 
 The results in the top (bottom) panel corresponds to the linear (full) fit. The black dots correspond to $\chi^2=0$.
  The second minimum (non-SM best fit point) of the full fit is at $\left\{ \ (c_{Hq}^{(1)}+c_{Hq}^{(3)}) ,\, c_{Hb} ,\, c_{bB} ,\,  c_{bW} \right\} = \left\{ -2.57,0.139,-2.70, -17.6 \right\} \times 10^{-2}$. For convenience, $\Lambda$ is set to 1\,TeV for all the coefficients here.
  }
	\label{fig:zpole240chq}
\end{figure}

\begin{figure}[htb]
	\centering

    \subfigure[~keeping only the linear dependence of Wilson coefficients]{
		
		\includegraphics[width=0.48\textwidth]{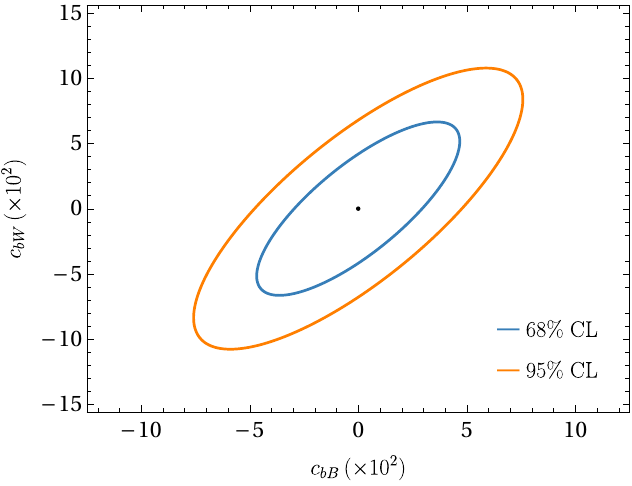}
	\hspace{0.2cm}
		\includegraphics[width=0.48\textwidth]{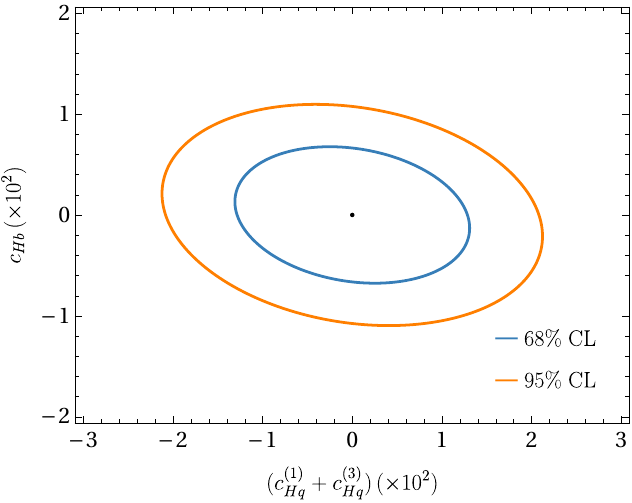}}
	
	\subfigure[~keeping the full dependence of Wilson coefficients]{
		
		\includegraphics[width=0.48\textwidth]{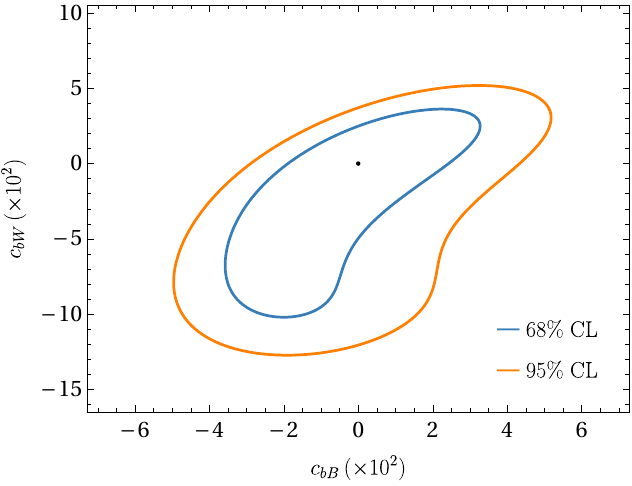}
	\hspace{0.2cm}
		\includegraphics[width=0.48\textwidth]{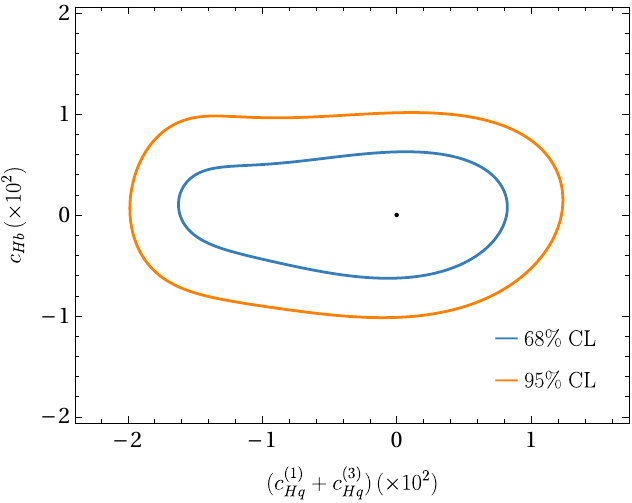}}
	
	\caption{%
    Same as \autoref{fig:zpole240chq} but also includes the 360\,GeV measurements.  The second minimum in the full fit is lifted. $\Lambda$ is also set to 1\,TeV here.
    }
	\label{fig:zpole240360chq}
\end{figure}

Here we present the results in terms of the dim-6 Wilson coefficients     $(c_{Hq}^{(1)}+c_{Hq}^{(3)})$ , $c_{Hb} $ ,  $ c_{bB}$  and  $c_{bW} $.  The $ 1\sigma $ bounds and correlations are summarized in \autoref{tab:fit2}, while the preferred regions are shown in \autoref{fig:zpole240chq} (for Z-pole + 240\,GeV) and \autoref{fig:zpole240360chq} (for Z-pole + 240\,GeV + 360\,GeV).  Note that the results here are the same as the ones in \autoref{sec:results} but only in a different basis.

\begin{table}[t]
	\centering
	\renewcommand\arraystretch{1.0}
	\vspace{0.5em}\centering
	\setlength{\tabcolsep}{0.5em}{
 
    \begin{tabular}{cccccc|c}
        \multicolumn{7}{c}{Z-pole + 240\,GeV } \\
        \hline 
                & Linear fit                             & \multicolumn{4}{c|}{Correlation $\rho$}                   & Full fit                            \\
                & $1 \sigma$ bound ($\times 10^{-2}$) & $(c_{Hq}^{(1)}+c_{Hq}^{(3)})$ & $c_{Hb}$ & $c_{bB}$ & $c_{bW}$ & $1 \sigma$ bound ($\times 10^{-2}$) \\ \hline
    $(c_{Hq}^{(1)}+c_{Hq}^{(3)})$ & $\pm 0.875$                          & 1               &                 &               &               &     [-2.94, 0.726]   \\
    $c_{Hb}$  & $\pm 0.45$                          &  -0.200           & 1               &               &               & [-0.439, 0.554]                \\
    $c_{bB}$   & $\pm 3.13 $                         & 0.814          &  -0.255          & 1             &               & [-5.10, 2.85]                 \\
    $c_{bW}$   & $\pm 4.45$                         & 0.975          &    -0.261        & 0.776         & 1             & [-20.57, 3.58]                \\ \hline \vspace{0cm}
    \end{tabular}    }
    \setlength{\tabcolsep}{0.5em}{

    \begin{tabular}{cccccc|c}
        \multicolumn{7}{c}{Z-pole + 240\,GeV + 360\,GeV } \\
        \hline 
                & Linear fit                             & \multicolumn{4}{c|}{Correlation $\rho$}                   & Full fit                            \\
                & $1 \sigma$ bound ($\times 10^{-2}$) & $(c_{Hq}^{(1)}+c_{Hq}^{(3)})$ & $c_{Hb}$ & $c_{dB}$ & $c_{dW}$ & $1 \sigma$ bound ($\times 10^{-2}$) \\ \hline
    $(c_{Hq}^{(1)}+c_{Hq}^{(3)})$ & $\pm 0.867$                          & 1               &                 &               &               &[-0.950, 0.554]                  \\
    $c_{Hb}$ & $\pm 0.448$                          &  -0.193           & 1               &               &               &         [-0.396, 0.396]    \\
    $c_{bB}$   & $\pm 3.10$                         & 0.814          & -0.249          & 1             &               & [-2.40, 2.16]                 \\
    $c_{bW}$   & $\pm 4.41$                         & 0.974         &  -0.254        & 0.777         & 1             & [-5.55, 2.55]                \\ \hline
    \end{tabular}   }
    	\caption[results]{1$ \sigma $ bound and correlations of $(c_{Hq}^{(1)}+c_{Hq}^{(3)})$ , $c_{Hb} $ , $ c_{bB}$ , $c_{bW} $ with expected measurements of EW observables at CEPC. For convenience, $\Lambda$ is set to 1\,TeV for all the coefficients here.
     }
     \label{tab:fit2}
\end{table}

%

%%%%%%%%%%%%%%%%%%%%%%%%%%%%%%%%%%%%
\section{Results by using running mass $m_b(m_Z)=2.86\GeV$}
\label{app:runningmass}

In \autoref{tab:fit1-run} and \ref{tab:fit2-run}, we present the fit results with combined Z-pole, 240\,GeV and 360\,GeV measurements, which uses the bottom mass value $m_b(m_Z)=2.86\GeV$ (in the $\overline{\rm MS}$ scheme with the renormalization scale $\mu=m_Z$).  They should be compared with the (bottom panel of) \autoref{tab:fit1} and \ref{tab:fit2}, which use $m_b(m_b)=4.18\GeV$.
For the linear fit, we expect the bounds for the dipole coefficients to be worsen by a factor of $m_b(m_b)/m_b(m_Z)\approx 1.5$, which is indeed the case for $c_{bB}$ and $c_{bW}$ in \autoref{tab:fit2-run}. 
For $\kappa_{bZ}$ and $\kappa_{bA}$ in \autoref{tab:fit1-run}, however, there appears to be little change in the linear fit result.  This is merely due to the parameterization we use in \autoref{eq:La}, where $\kappa_{bZ}$ and $\kappa_{bA}$ are normalized by a factor of $m_b$. For the full fit, the change in the results are somewhat less predictive due to the nontrivial interplay between linear and quadratic terms.  Nevertheless, the overall results for $c_{bB}$ and $c_{bW}$ are still worse for a smaller $m_b$ as expected. 

\begin{table}[htb]
    \centering
	\renewcommand\arraystretch{1.0}
	\vspace{0.5em}\centering
	\setlength{\tabcolsep}{0.5em}{
    \begin{tabular}{cccccc|c}
        \multicolumn{7}{c}{Z-pole + 240\,GeV + 360\,GeV  (with $m_b=2.86$\,GeV) } \\
        \hline 
               & Linear fit                             & \multicolumn{4}{c|}{ Correlation $\rho$}                   & Full fit                            \\
                & $1 \sigma$ bound ($\times 10^{-4}$) & $\delta g_{Lb}$ & $\delta g_{Rb}$ & $\kappa_{bZ}$ & $\kappa_{bA}$ & $1 \sigma$ bound ($\times 10^{-4}$) \\ \hline
    $\delta g_{Lb}$ & $\pm 2.63$                          & 1               &                 &               &               & [-2.64, 1.52]                   \\
    $\delta g_{Rb}$ & $\pm 1.34$                          & -0.188           & 1               &               &               & [-1.20, 1.20]                   \\
    $\kappa_{bZ}$   & $\pm 0.373$                         & 0.975           & -0.260           & 1             &               & [-0.390, 0.195]                \\
    $\kappa_{bA}$   & $\pm 0.125$                         & 0.115           & -0.090          & 0.205         & 1             & [-0.072, 0.186]                 \\ \hline
    \end{tabular}   }   
    	\caption[results]{
        1$ \sigma $ bound and correlations of $ \delta g_{Lb} $, $ \delta g_{Rb} $, $ \kappa_{bZ} $, $ \kappa_{bA} $ with expected measurements of EW observables at CEPC, using a bottom mass of $m_b(m_Z)=2.86\GeV$ instead of the $m_b(m_b)=4.18\GeV$ in \autoref{tab:fit1}.  
     }
    \label{tab:fit1-run}
\end{table}

\begin{table}[t]
	\centering
	\renewcommand\arraystretch{1.0}
	\vspace{0.5em}\centering
    \setlength{\tabcolsep}{0.5em}{

    \begin{tabular}{cccccc|c}
        \multicolumn{7}{c}{Z-pole + 240\,GeV + 360\,GeV (with $m_b=2.86$\,GeV) } \\
        \hline 
                & Linear fit                             & \multicolumn{4}{c|}{Correlation $\rho$}                   & Full fit                            \\
                & $1 \sigma$ bound ($\times 10^{-2}$) & $(c_{Hq}^{(1)}+c_{Hq}^{(3)})$ & $c_{Hb}$ & $c_{dB}$ & $c_{dW}$ & $1 \sigma$ bound ($\times 10^{-2}$) \\ \hline
    $(c_{Hq}^{(1)}+c_{Hq}^{(3)})$ & $\pm 0.869$                          & 1               &                 &               &               &[-0.871, 0.501]                  \\
    $c_{Hb}$ & $\pm 0.444$                          &  -0.188           & 1               &               &               &         [-0.396, 0.396]    \\
    $c_{bB}$   & $\pm 4.54$                         & 0.815          & -0.246          & 1             &               & [-3.12, 3.12]                 \\
    $c_{bW}$   & $\pm 6.46$                         & 0.974         &  -0.249        & 0.778         & 1             & [-8.16, 3.23]                \\ \hline
    \end{tabular}   }
    	\caption[results]{
        1$ \sigma $ bound and correlations of $(c_{Hq}^{(1)}+c_{Hq}^{(3)})$ , $c_{Hb} $ , $ c_{bB}$ , $c_{bW} $ with expected measurements of EW observables at CEPC, using a bottom mass of $m_b(m_Z)=2.86\GeV$ instead of the $m_b(m_b)=4.18\GeV$ in \autoref{tab:fit2}.  For convenience, $\Lambda$ is set to 1\,TeV for all the coefficients here.
     }
     \label{tab:fit2-run}
\end{table}

%%%%%%%%%%%%%%%%%%%%%%%%%%%%%%%%%%%%

%references%%%%%%%%%%%%%%%%%%%%%%%%%
\bibliographystyle{JHEP}
\bibliography{paper.bib}

%%%%%%%%%%%%%%%%%%%%%%%%%%%%%%%

\end{document}